\documentstyle[aps,prb]{revtex}
\begin{document}

\input epsf.sty
\twocolumn[\hsize\textwidth\columnwidth\hsize\csname %
@twocolumnfalse\endcsname

\draft

\widetext

\title{Absence of magnetic field effect on static magnetic
order in electron-doped superconductor Nd$_{1.86}$Ce$_{0.14}$CuO$_4$}
\author{M. Matsuda and S. Katano}
\address{
Advanced Science Research Center,
Japan Atomic Energy Research Institute, Tokai, Ibaraki 319-1195, Japan}
\author{T. Uefuji, M. Fujita, and K. Yamada}
\address{
Institute for Chemical Research, Kyoto University, Gokasho, Uji
610-0011, Japan}

\date{\today}
\maketitle
\begin{abstract}
Neutron-scattering experiments were performed to study
the magnetic field effect on the electron-doped cuprate
superconductor Nd$_{1.86}$Ce$_{0.14}$CuO$_4$,
which shows the coexistence of magnetic order and
superconductivity.
The ($\frac{1}{2}\frac{3}{2}$0) magnetic Bragg intensity,
which originates from the order of
both the Cu and Nd moments at low temperatures,
shows no magnetic field dependence when the field is applied
perpendicular to the CuO$_{2}$ plane up to 10 T above
the upper critical field.
This result is significantly different from that reported
for the hole-doped cuprate superconductors,
in which the quasi-static magnetic order is noticeably enhanced
under a magnetic field.
\end{abstract}

\phantom{.}
]
\narrowtext

Extensive neutron-scattering studies have been performed
on high-$T\rm_c$ superconductors in order to clarify
the interplay between the superconductivity and magnetism.
In particular, in the hole-doped cuprate superconductor
La$_{2-x}$Sr$_x$CuO$_4$ and related systems,
static and dynamic properties of spin correlations have been studied
in considerable detail.~\cite{kastner,cheong,yamada0,tranquada,fujita0}
A remarkable feature in the superconducting phase is
that static and low-energy spin correlations are incommensurate
and the magnetic peaks are found at
($\frac{1}{2}$,$\frac{1}{2}\pm\delta$) and
($\frac{1}{2}\pm\delta$,$\frac{1}{2}$).~\cite{cheong,yamada0}
In the optimally doped region, there exists an excitation gap and
low-energy excitations are suppressed.~\cite{yamada1}
On the other hand, in the region where hole concentration
is $\sim \frac{1}{8}$, elastic incommensurate peaks,
originating from both the spin density wave and charge density
wave, are observed distinctly,
suggesting the stripe model.~\cite{tranquada}
In this underdoped region, the coexistence of magnetic order and
superconductivity is implied.~\cite{tranquada2,niedermayer}

In the electron-doped cuprate superconductor, however,
the number of neutron-scattering studies is rather limited, probably
because a large single crystal is difficult to grow.
Yamada $et$ $al$. reported that the superconducting
Nd$_{1.85}$Ce$_{0.15}$CuO$_4$ (superconducting transition
temperature $T_c\sim$18 K) shows a broad
magnetic excitation peak at the commensurate position
($\frac{1}{2}$,$\frac{1}{2}$).~\cite{yamada}
It is also found that an excitation gap exists around 4.5 meV.
Thus, both hole- and electron-doped cuprate superconductors show
the gap behavior in magnetic excitations although
the magnetic correlations are incommensurate and commensurate in
hole- and electron-doped cuprate superconductors, respectively.
The coexistence of magnetic order and superconductivity is also
suggested in the electron-doped system.~\cite{uefuji2,uefuji1,fujita}

Neutron-scattering under a magnetic field is one of
the important techniques that can be used to study the interplay
between magnetism and superconductivity.
The magnetic field effect has been studied in superconducting
La$_{2-x}$Sr$_x$CuO$_4$ ($x$=0.10 and 0.12) and
La$_2$CuO$_{4+y}$.
These investigations showed that the static parallel
stripe order is enhanced under a magnetic field perpendicular to the
CuO$_2$ planes.~\cite{katano,lake,lee}
The enhancement of the elastic magnetic
intensity is ascribed to the vortices which stabilize the static
magnetic order in a larger region than
the vortex cores.~\cite{katano,lake,lee}
Theoretical studies have also been performed intensively on the static
magnetic ordering induced near the vortex cores, which is consistent
with the experiments.~\cite{arovas,ogata,ichioka,zhu,hu,dhlee,franz,zhang,chen}

In the case of the electron-doped cuprate superconductor
Nd$_{2-x}$Ce$_x$CuO$_4$, magnetic field studies have been performed
only for undoped Nd$_2$CuO$_4$,~\cite{skanthakumar}
to the best of our knowledge.
The main purpose of that study
was to determine whether the magnetic structure is collinear or
noncollinear. The magnetic field was applied in the
CuO$_2$ plane and the magnetic structure was found to be noncollinear.
In the present study, we examined the magnetic field effect of
the static magnetic
correlations in Nd$_{1.86}$Ce$_{0.14}$CuO$_4$ ($T_c\sim$25 K).
Since the coherence length is $\sim$100 \AA\
in the electron-doped system,~\cite{hidaka}
which is several times larger than that in La$_{2-x}$Sr$_x$CuO$_4$,
a large magnetic field effect can be expected. Furthermore, the upper
critical field $H_{c2}$ is less than 10 T in the electron-doped system
so that normal-state properties can easily be studied.
It is found that the elastic magnetic peak is magnetic field
independent up to 10 T above $H_{c2}$, suggesting that the interplay
between magnetic order and superconductivity in this system is
considerably different from that in the hole-doped system.

The single crystal of Nd$_{1.86}$Ce$_{0.14}$CuO$_4$
was grown by the traveling solvent floating-zone method.
The crystals were annealed in an Ar atmosphere at 920 $^\circ$C
for 12 h. $T_c$ is $\sim$25 K
as determined from a susceptibility measurement, and is shown in Fig. 1.
From the data, the superconducting property is considered to be that
of bulk in nature.
The crystal used in this study is the one that was used
in the previous study.~\cite{uefuji1}
The Ce concentration dependence of magnetic and
superconducting properties shows a systematic change in the transition
temperatures, indicating that the doped electrons are homogeneously
distributed.~\cite{uefuji2,uefuji1}
For $x$=0.14, which was used in this study, similar volume of
a magnetic ordered phase and superconducting phase
coexist.~\cite{uefuji2}
It is likely that a slightly inhomogeneous distribution of
electrons causes two phases that are spatially separated.
This phase separation behavior is basically similar to that
in La$_{2-x}$Sr$_x$CuO$_4$.

The neutron-scattering experiments were carried out on
the three-axis spectrometer TAS2 installed in the
guide hall of JRR-3M at the Japan Atomic Energy Research Institute.
The typical horizontal collimator sequence was
guide-20$'$-S-20$'$-80$'$ with a fixed
incident neutron energy of $E\rm_i$=13.7 meV.
Contamination from higher order beams was effectively eliminated
using pyrolytic graphite filters.
The single crystal was oriented in the $(HK0)$ scattering plane.
The neutron-scattering experiments under magnetic fields were
performed up to 10 T using a new type of split-pair superconducting 
magnet cooled by cryocoolers.
The field was applied vertically to the 
scattering plane.
\begin{figure}
\centerline{\epsfxsize=3.2in\epsfbox{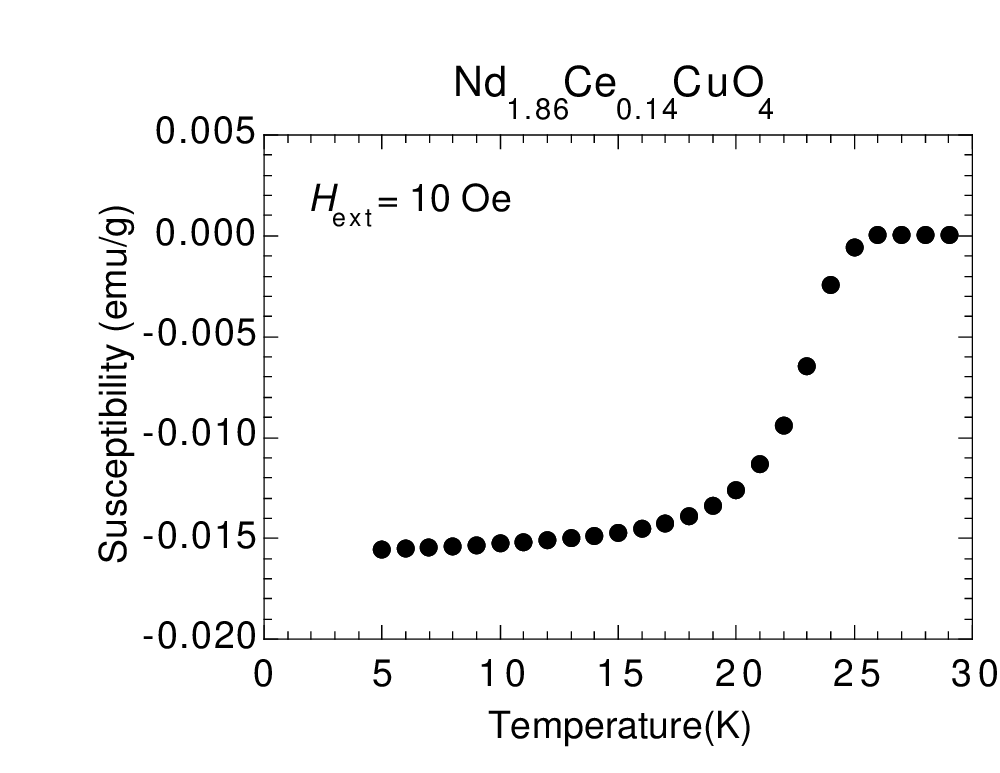}}
\caption{Temperature dependence of magnetic susceptibility
under a zero-field-cooled condition
with the magnetic field of 10 Oe for Nd$_{1.86}$Ce$_{0.14}$CuO$_4$.}
\label{fig1}
\end{figure}

In the $(HK0)$ scattering zone, magnetic Bragg peaks are observed
at ($\frac{1}{2}+m,\frac{1}{2}+n$,0), where $m$ and $n$ are integers,
except at $(\pm\frac{1}{2}\pm m,\pm\frac{1}{2}\pm m,0)$ and
$(\pm\frac{1}{2}\pm m,\mp\frac{1}{2}\mp m,0)$.
It is also reported that superlattice peaks, which originate from a
superstructure caused by the heat treatment and are almost temperature
independent, are superposed on these magnetic
peak positions.~\cite{kurahashi}
Figure 2 shows the temperature dependence of the
($\frac{1}{2}\frac{3}{2}$0) Bragg intensity under
zero magnetic field.
The intensity at ($\frac{1}{2}+m,\frac{1}{2}+n$,0)
is described as

\begin{eqnarray}
I=C\left\{M{\rm_{Cu}}(T)f{\rm_{Cu}}
+2M{\rm_{Nd}}(T)f{\rm_{Nd}}\right\}^2+I{\rm_{lattice}}
\label{int}
\end{eqnarray}
where $C$ is a constant, $I\rm_{lattice}$ is the superlattice
scattering intensity, and $M(T)$ and $f$ are ordered staggered moments
and form factors for the Cu$^{2+}$ and Nd$^{3+}$ ions, respectively.
Above 100 K, this reflection almost originates from the
superstructure. With decreasing temperature, the Cu and Nd moments
order gradually and a contribution from the order of the Cu moments
becomes comparable to that from the superstructure around 50 K.
Below $\sim$20 K the order of the Nd moments develops rapidly
so that most of the scattering intensity originates
from the magnetic order and the contribution of the Nd moments
becomes comparable to that of the Cu moments.
It was reported that $M\rm_{Cu}$(10 K)$\sim0.1\mu_B$ and
$M\rm_{Nd}$(10 K)$\sim0.05\mu_B$ in Nd$_{1.86}$Ce$_{0.14}$CuO$_4$
if the moments are assumed to be homogeneously distributed.~\cite{uefuji1}

Figure 3 shows the magnetic field dependence of
the neutron elastic intensity at ($\frac{1}{2}$,$\frac{3}{2}$,0) in
Nd$_{1.86}$Ce$_{0.14}$CuO$_4$. The magnetic field is applied
perpendicular to the CuO$_2$ plane.
The magnetic peak width is slightly broader than the instrumental
resolution~\cite{uefuji1} although the superlattice peak is almost
resolution-limited in the $(HK0)$ plane.~\cite{kurahashi}
At 75 K, where most of the scattering intensity comes from the structural
distortion, there is no magnetic field effect up to 10 T,
which is reasonable.
At 45 K above $T\rm_c$, where about one third of the 
intensity comes from the static magnetic order, mostly of
the Cu moments,~\cite{moment} the magnetic field effect is still missing.
Finally, at 15 K below $T\rm_c$, where about 80\% of the intensity is
magnetic in origin and the Cu and Nd contributions are
comparable,~\cite{moment} there is almost no magnetic field dependence
even at 10 T, which is above $H_{c2}$.
This result is significantly different from that reported
for the hole-doped system, in which the quasi-static magnetic order
is enhanced under a magnetic field.~\cite{katano,lake,lee}
\begin{figure}
\centerline{\epsfxsize=2.6in\epsfbox{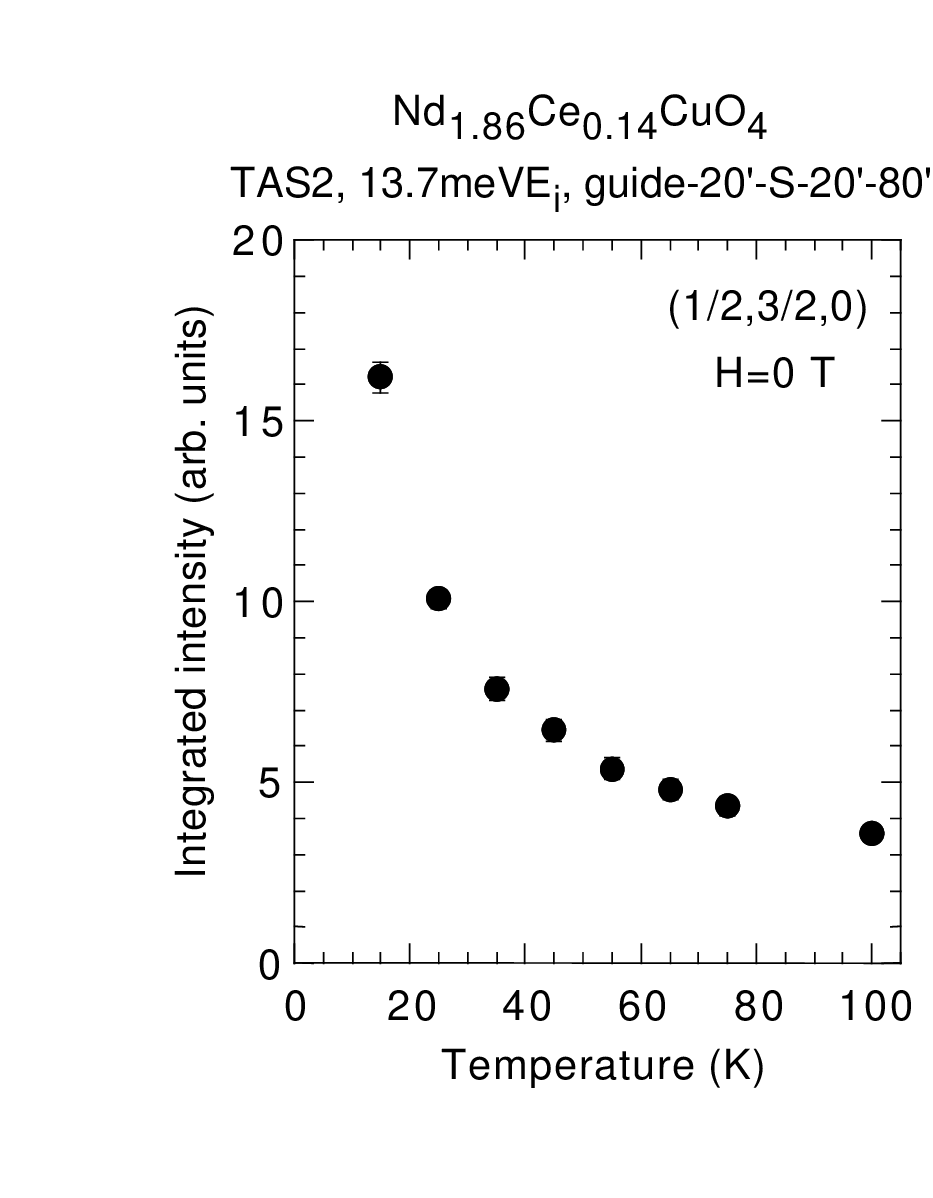}}
\caption{Temperature dependence of the ($\frac{1}{2}\frac{3}{2}$0)
Bragg peak under zero magnetic field in Nd$_{1.86}$Ce$_{0.14}$CuO$_4$.}
\label{fig2}
\end{figure}

As mentioned at the beginning for hole-doped La$_{2-x}$Sr$_x$CuO$_4$,
it is theoretically predicted
that the static magnetic order is stabilized and enhanced around
the vortex cores with an application of magnetic
field,~\cite{arovas,ogata,ichioka,zhu,hu,dhlee,franz,zhang,chen}
indicating that dynamic spin fluctuations in the superconducting phase
can be easily pinned by the vortices.
If such a strong pinning effect also exists in the electron-doped system,
the magnetic field should enhance the static magnetic order.
Since almost the same volume of magnetic and superconducting
phases coexist in Nd$_{1.86}$Ce$_{0.14}$CuO$_4$,~\cite{uefuji2}
enhancement of the elastic magnetic intensity is expected to be well
observable.

\begin{figure}
\centerline{\epsfxsize=2.5in\epsfbox{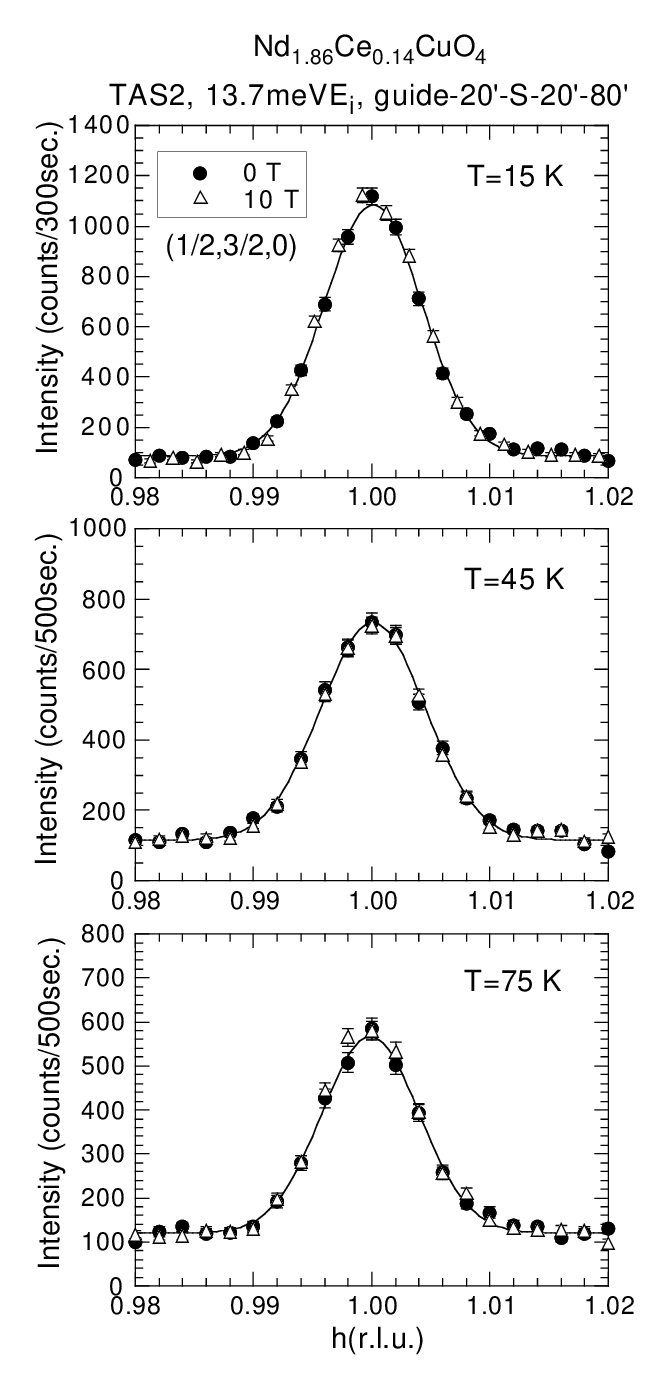}}
\caption{Neutron elastic intensity around the
commensurate position ($\frac{1}{2},\frac{3}{2}$,0)
in Nd$_{1.86}$Ce$_{0.14}$CuO$_4$
under magnetic fields $H$=0 and 10 T and at $T$=15, 45, and 75 K.
Magnetic field is applied perpendicular to the CuO$_2$ plane.
The solid lines are the results of fits to
a Gaussian function for the zero field data.
}
\label{fig3}
\end{figure}
A puzzling question is how do the magnetically ordered phase and
superconducting phase coexist in the electron-doped system.
We mentioned that the two phase behavior probably originates from
the phase separation of the doped carriers, which is also probable
in La$_{2-x}$Sr$_x$CuO$_4$.
Although the correlation length of the magnetic ordered phase
in the CuO$_2$ plane is similar in the both systems ($\sim$100 \AA),
the deference between the two systems is
the correlation length perpendicular to the CuO$_2$ plane.
In nearly optimum-doped region of La$_{2-x}$Sr$_x$CuO$_4$,
magnetic correlations are almost two-dimensional.
On the other hand, the correlation length in Nd$_{1.86}$Ce$_{0.14}$CuO$_4$
is estimated to be about 100 \AA,\cite{uefuji1} which is fairy large.
Therefore, the magnetic ordered state is expect to be robust
in Nd$_{1.86}$Ce$_{0.14}$CuO$_4$ against magnetic fields,
which is consistent with the experimental results.
Even in this case, however, a static magnetic order might appear in the
superconducting region
when the magnetic field exceeds $H_{c2}$ and the superconducting region
turns to be in the normal-state.
Therefore, the absence of the magnetic field effect is surprising.

A probable scenario for the absence of the magnetic field effect
would be as follows.
The superconducting phase has an excitation gap
as reported in Nd$_{1.85}$Ce$_{0.15}$CuO$_4$~\cite{yamada}
and the gap does not close even above $H_{c2}$
so that a quasi-elastic component still does not appear.
This behavior is similar to that observed in optimally doped
La$_{2-x}$Sr$_x$CuO$_4$,~\cite{lake0} in which an in-gap state develops
but the gap still remains under a magnetic field,
although the applied field is much smaller than $H_{c2}$.
It is also possible that the superconducting phase in the
Nd$_{2-x}$Ce$_x$CuO$_4$ system lies in the overdoped phase,
in which the magnetic fluctuations are shorter ranged \cite{yamada0}
and thus a long-range magnetic order does not appear easily.
In order to clarify this in detail, the magnetic field dependence of
the excitation spectra should be measured. We plan to perform
neutron inelastic scattering experiments under a magnetic field in future.

In summary, our neutron-scattering experiments under a magnetic field
in the electron-doped cuprate
superconductor Nd$_{1.86}$Ce$_{0.14}$CuO$_4$ demonstrate that the
static magnetic order of both the Cu and Nd moments shows no
magnetic field dependence up to 10 T which is above $H_{c2}$.
This is in sharp contrast to the static magnetic order
in the hole-doped system.

\section*{Acknowledgments}
We would like to thank S.-H. Lee
for stimulating discussions and Y. Shimojo for technical assistance.
This study was supported in part
by a Grant-in-Aid for Scientific Research from the Japanese Ministry
of Education, Culture, Sports, Science, and Technology.

\end{document}